\documentclass{jltp}
\usepackage{graphicx} 

\title{Iordanskii and Lifshitz-Pitaevskii Forces in the Two-Fluid Model}

\author{M. R. Geller$^1$, D. J. Thouless,$^2$ S. W. Rhee,$^2$ and W. F. 
Vinen$^3$}

\address{$^1$Department of Physics, University of Georgia, Athens GA 30602, USA\\
$^2$Department of Physics, University of Washington, Seattle, WA 98195, USA\\
$^3$School of Physics, University of Birmingham, Birmingham B15 2TT, UK}

\runninghead{M. R. Geller {\it et al.}}
{Iordanskii and Lifshitz-Pitaevskii Forces}

\begin{document}
\maketitle
  
\begin{abstract}
It has been known since the pioneering work of Onsager and Feynman 
that the statistical mechanics and dynamics of vortices play an 
essential role in the behavior of superfluids and superconductors. 
However, the theory of vortices in quantum fluids remains in a most 
unsatisfactory state, with many conflicting results in the literature.
In this paper we review the theory of Thouless, Ao and Niu, which 
gives an expression for the total transverse force acting on a 
quantized vortex that is in apparent disagreement with the work of 
Iordanskii and of Lifshitz and Pitaevskii. In particular, no 
transverse force proportional to the asymptotic normal fluid velocity 
was found. We use two-fluid hydrodynamics to study this discrepancy.

\end{abstract}

\section{A HIERARCHY OF LENGTH SCALES}

A comprehensive theory of vortex dynamics in quantum fluids must 
address phenomena occurring at three different length scales. At the 
smallest, most microscopic scale, a fully quantum mechanical treatment
of a vortex, including its internal structure, interaction with 
elementary excitations such as phonons and rotons, and interaction 
with disorder, is required

At the next---what we shall refer to as intermediate---length scale, 
one would like to regard a vortex as a classical object, subject to a 
classical equation of motion. Of course, there is no guarantee that we
can do this, and, indeed, certain pathologies result from our 
insistence in doing so, but experience has shown that this classical 
picture is extremely successful.\cite{Donnelly}

There are different ways to formulate the classical approach. If we 
let  ${\bf R}(t)$ denote the position in the $xy$ plane of the center 
of an isolated straight vortex line, say, as a function of time, we 
could hypothesize a phenomenological equation of motion of the form
\begin{equation}
M \, {d^2 {\bf R}\over dt^2} = -\eta \, {d {\bf R} \over dt} - \gamma 
\, {d {\bf R} \over dt} \times {\bf e}_z + {\bf f}_{\rm p}({\bf R}) 
+ {\bf f}_{\rm d}({\bf R}).
\label{equation of motion}
\end{equation}
Here we have taken the circulation vector ${\bf K}$ of the vortex, a 
vector parallel to the vortex with a magnitude equal to the 
circulation, to be along the $z$ direction. The first two terms on 
the right-hand-side of (\ref{equation of motion}) are to include 
all forces linear in the vortex velocity. The coefficients $M$, 
$\eta$, and $\gamma$, which describe the vortex effective mass per 
unit length, viscous damping force per unit length, and nondissipative
transverse force per unit length, respectively, are to be determined
from a microscopic theory, as are the ``pinning'' and ``driving''
forces per unit length ${\bf f}_{\rm p}$ and ${\bf f}_{\rm d}$.
The latter may depend on the normal and superfluid densities and
velocities, and therefore on the vortex position ${\bf R}$, but do not
depend on the vortex velocity. In a superfluid, ${\bf f}_{\rm p}$ 
might describe the force exerted by an externally imposed 
wire, and ${\bf f}_{\rm d}$ would include the 
superfluid-velocity-dependent part of the Magnus force (see below)
and possibly other vortex-velocity-independent contributions.

At the most macroscopic scale one needs to understand how the forces 
arising at the intermediate scale act to determine the bulk, 
experimentally observable properties of quantum fluids. The length 
scale that usually defines this regime is the characteristic 
inter-vortex distance, and one is interested in coarse-grained 
properties of the quantum fluid at scales larger than that distance. 
In superfluids, this is the regime where the concept of mutual 
friction applies. Mutual friction refers to a momentum transfer, with
both longitudinal and transverse components, between the normal and 
superfluid parts of a quantum fluid.\cite{Donnelly} Although such an 
interaction is 
absent in the two-fluid model itself, the presence of vortices mediate
a bulk momentum exchange. Similarly, in superconductors one needs to 
understand how the Lorentz force, pinning forces, and other 
intermediate-scale forces conspire to determine, say, the Hall 
effect in the mixed state, which depends on the collective motion 
of a macroscopic number of vortices.\cite{Blatter etal}

Our recent work has focused mostly on the intermediate length-scale 
regime, namely, the determination of the various forces that act on 
vortices. After a brief general review of that work we shall discuss 
our new results on the problem of the Iordanskii and 
Lifshitz-Pitaevskii forces, which have been subjects of considerable 
controversy.

\section{THE TAN THEORY}

The Thouless-Ao-Niu (TAN) theory\cite{Thouless Ao and Niu} and its 
generalization to 
ultraclean type-II superconductors\cite{Geller Wexler and Thouless} 
start with the
appropriate exact many-body Hamiltonian and include a pinning 
potential centered at ${\bf R}$, which is also the position of the 
vortex. The vortex is then dragged with a velocity ${\bf V}$, and a 
transverse force ${\bf f}_\perp = {\bf e}_z \times {\bf V} \, \oint 
d{\bf l} \cdot {\bf j}_{\rm c}$ is found. Here ${\bf j}_{\rm c}$ is 
the canonical momentum density, and the line integral is taken around 
a large circle enclosing the vortex. Evidently, the transverse force 
is independent of the detailed microscopic structure of the vortex 
and its interaction with the normal fluid.

In a neutral Bose or Fermi superfluid, ${\bf j}_{\rm c} = \rho_{\rm s}
{\bf v}_{\rm s} + \rho_{\rm n} {\bf v}_{\rm n},$ and therefore  
\begin{equation}
\oint d{\bf l} \cdot {\bf j}_{\rm c} = \rho_{\rm s} K_{\rm s} + 
\rho_{\rm n} K_{\rm n}.
\label{SF case}
\end{equation}
In the normal fluid component, viscosity causes
vortex-like circulation to diffuse away to the outer boundary, so $K_{\rm n}$
vanishes in equilibrium, and
\begin{equation}
{\bf f}_\perp = \rho_{\rm s} {\bf K}_{\rm s} \times {\bf V}. 
\label{SF transverse force}
\end{equation}

In a charged superfluid or superconductor, however, the Hamiltonian 
contains a current-current interaction term that modifies the velocity
operator. The canonical momentum density can be expressed in terms 
of the physical, gauge-invariant momentum density ${\bf j} = 
{\bf j}_{\rm c} + {\textstyle{e \over c}} n {\bf A}$, resulting in
\begin{equation}
\oint d{\bf l} \cdot {\bf j}_{\rm c} = \oint d{\bf l} \cdot \big( 
{\bf j} - {\textstyle{e \over c}} n {\bf A} \big) = \rho \, \Phi_0 ,
\ \ \ \ \ \ \ \Phi_0 \equiv hc/2e.
\label{SC case}
\end{equation}
The second equality in (\ref{SC case}) follows from the Meissner 
effect, which causes ${\bf j}$ to vanish at large distances, and from 
flux quantization. Therefore,
\begin{equation}
{\bf f}_\perp = \rho {\bf K}_{\rm s} \times {\bf V},
\end{equation}
where $\rho = \rho_{\rm s} + \rho_{\rm n}$ is the total mass density
of the fluid.

\section{A HIERARCHY OF CONTROVERSIES}

We turn now to a brief discussion of some of the controversies in 
the theory of intermediate-scale vortex dynamics, focusing on
transverse forces, and organized according to the complexity of the 
quantum fluid in question.

It seems appropriate to start by recalling a result of classical 
hydrodynamics, the Magnus force: When a vortex with circulation 
${\bf K}$ is dragged with velocity ${\bf V}$ through an ideal fluid  
of mass density $\rho$, a transverse force per unit length 
${\bf f}_\perp = \rho \, {\bf K} \times {\bf V}$ acts on the object 
doing the dragging. The force is similar to the lift force on an 
airplane wing. It is clear from Galilean invariance that if the fluid 
far from the vortex is not at rest, but has a velocity ${\bf v}$, then
the force is instead
\begin{equation}
{\bf f}_\perp = \rho \, {\bf K} \times ({\bf V}-{\bf v}).
\label{magnus force}
\end{equation}

By analogy, it is reasonable in a neutral Bose superfluid at zero 
temperature to expect that ${\bf f}_\perp = \rho_{\rm s} \, 
{\bf K}_{\rm s} \times ({\bf V}-{\bf v}_{\rm s}),$
where $K_{\rm s}$ is the quantized circulation. Of course, writing
$\rho_{\rm s}$ here instead of $\rho$ is arbitrary, because these are 
equal at zero temperature.

The controversy begins in the finite-temperature neutral Bose 
superfluid, because Galilean invariance allows for a transverse force 
of the form
\begin{equation}
{\bf f}_\perp = a \, {\bf K}_{\rm s} \times ({\bf V}-{\bf v}_{\rm s})
+ b \, {\bf K}_{\rm s} \times ({\bf V}-{\bf v}_{\rm n}),
\label{general SF magnus force}
\end{equation}
where $a$ and $b$ are parameters. The first term in (\ref{general SF 
magnus force}) has a simple classical interpretation: It describes a 
Magnus-type force originating from the superfluid component of the 
fluid (remembering that it is the superfluid that is 
circulating). Thus classical reasoning would suggest that $a = 
\rho_{\rm s}$, and Wexler has recently given an elegant proof of 
this.\cite{Wexler} To our knowledge there have been no criticisms of 
Wexler's theory.

From this classical point-of-view, again keeping in mind that it is 
the superfluid that is circulating here, the second term in 
(\ref{general SF magnus force}) would be of a nonhydrodynamic origin. 
If present, it would describe a transverse interaction between the 
vortex and the excitations---phonons and rotons---of the fluid. 

Iordanskii\cite{Iordanskii} predicted just such an interaction with 
phonons, and Lifshitz and Pitaevskii\cite{Lifshitz and Pitaevskii} 
(following earlier work by Hall and Vinen\cite{Hall and Vinen})
predicted one with rotons. Note, however, that in these works
${\bf v}_{\rm s}$ and ${\bf v}_{\rm n}$ refer to flow velocities
{\it near} the vortex line, whereas our quantities are
asymptotic values. But the TAN 
result (\ref{SF transverse force}) implies $a + b = \rho_{\rm s}$. 
When combined with Wexler's theory, this implies $b=0$, i.e., that 
there are no transverse Iordanskii and Lifshitz-Pitaevskii forces. This
apparent discrepancy has motivated us to understand better the interaction 
between a quantized vortex and the normal fluid in a neutral Bose 
system, and to calculate the coefficient $b$ in (\ref{general SF 
magnus force}) directly. We shall return to this direct calculation 
below in Section \ref{two-fluid}

The next controversy concerns the transverse force in a neutral
Fermi superfluid, also described in the TAN theory. According to
Kopnin and Kravtsov,\cite{Kopnin and Kravtsov} low-energy 
quasiparticles in the vortex core also contribute to the transverse 
force linear in ${\bf V}$, a contribution not found by TAN.

Vortices in superconductors inherit all of the above controversies and
have additional complexity of their own.\cite{Sonin,Ao and Zhu} We will 
not have space to discuss them further.

\section{VORTEX DYNAMICS IN THE TWO-FLUID MODEL} 
\label{two-fluid}

Two ingredients are needed for a microscopic theory of the coefficient
$b$ in Eqn.~(\ref{general SF magnus force}). First, one has to solve 
a vortex-excitation scattering problem. 
It is probably correct to use the Gross-Pitaevskii equation to do 
this, even though mean field theory is not expected to hold inside 
the vortex, because the scattering potential is long-ranged. In fact,
the potential is sufficiently long-ranged, and the forward scattering
sufficiently singular, that there are several mathematically 
incorrect scattering calculations present in the 
literature.\cite{Wexler and Thouless} 

In the scale of lengths greater than the size of the vortex core and less
than the mean free path for scattering of excitations by one another, if
there is such a range, the dynamics of excitations moving in the slowly
varying background of the rotating superfluid is well defined, and in
this region the traditional arguments\cite{Iordanskii,Lifshitz and 
Pitaevskii,Nielsen and Hedegard}
give a transverse force on the excitations equal to $\rho_{\rm n} \, 
{\bf K}_{\rm s} \times({\bf V}-{\bf v}_{\rm n})$, where ${\bf v}_{\rm n}$ 
is the average normal
fluid velocity at a distance of the order of the mean free path from the
vortex core, as well as a dissipative longitudinal force.

In the region well beyond a mean free path from the vortex core, flow
velocities are varying slowly and two-fluid hydrodynamics, which incorporates 
the basic conservation laws, provides an accurate description.  In
particular, the two-fluid version of the Navier--Stokes equation gives
the force--momentum balance.  In this region we have studied how forces
generated within a mean free path of the vortex core can be transmitted
to large distances.\cite{Thouless Geller and Vinen}  
Within a linearized approximation to the two-fluid
model the force due to the superfluid flow relative to the vortex is just
the ordinary superfluid Magnus force $\rho_{\rm s} \, {\bf K}_{\rm s}
\times({\bf V}-{\bf v}_{\rm s})$, where ${\bf v}_{\rm s}$ is the asymptotic 
value of the superfluid
velocity. This is given for superfluids, as it is for ordinary fluids, by
the cross terms, between circulation and the superfluid flow, in the
momentum flow tensor and the Bernoulli pressure. There seems to be no
possible modification of this even for a fermionic superfluid, such as is
suggested by the work of Kopnin and Kravtsov,\cite{Kopnin and Kravtsov} 
if there is no bulk interaction with a stationary background. 

For the normal fluid contributions our considerations are quite similar
to those of Hall and Vinen.\cite{Hall and Vinen}
In the linear regime the force is transmitted by the viscous force (and
an accompanying pressure term), which is generated by a flow velocity in
the direction of the force that increases as the logarithm of the
distance from the vortex core. The asymptotic value of the normal fluid
velocity is given by a combination of the value ${\bf v}_{\rm n}$ close to the
core, and this logarithmically growing term.  The logarithmically growing
term has to be cut off at a distance $R_c$ which is determined by the
Reynolds number of the flow, by the spacing between vortices, or by the
diffusion length at the frequency of vortex oscillation.  In the limit
considered by TAN, with vanishingly small normal fluid velocity, and a
single vortex in an infinite medium, this cut-off goes to infinity, and
${\bf v}_{\rm n}$ is negligibly small in comparison, so the asymptotic 
velocity is 
determined by the logarithmic term, and the force is parallel to the normal 
fluid flow.

Realistically, the logarithm which the cut-off introduces is not
necessarily large compared with other parameters in the problem.  Even
when the Iordanskii force is not put in explicitly at the inner boundary,
there is some transverse force due to normal fluid flow, but this is
proportional to $[\ln (R_c/\lambda)]^{-2} \, v_{\rm n}$, where $\lambda$ is the
mean free path.  It is not clear how to match conditions close to the
core with conditions in the hydrodynamic region, but if it is assumed
that the Iordanskii force causes ${\bf v}_{\rm n}$ close to the core to 
match the
viscous force in the hydrodynamic region, and to be parallel to it, the
relative importance of the logarithmic term will depend on the ratio of
the kinematic viscosity $\eta/\rho_{\rm n}$ to the quantum of 
circulation---a large kinematic viscosity will diminish the importance of 
the logarithmic term in the flow velocity.

\section*{ACKNOWLEDGMENTS}
This work was supported in part by NSF grant No.~DMR-9813932, by a 
Research Innovation Award from the Research Corporation, and by a 
Sarah Moss Fellowship.

\end{document}